\RequirePackage{filecontents}

\RequirePackage{snapshot}
\documentclass[aps,prl,reprint,showpacs,secnumarabic,superscriptaddress]{preprint}

\def\Preprint{}
\renewcommand\documentclass[2][]{}
\let\origparagraph=\paragraph
\AtBeginDocument{%
  \let\paragraph=\origparagraph
  \let\section=\paragraph
  \let\ack=\acknowledgments
}
\makeatletter
\newcommand*{\br}{}
\newcommand*{\mr}{\colrule}
\renewenvironment{ruledtabular}{\def\array@default{v}%
  \let\tableft@skip@default\tableft@skip
  \let\tableft@skip\tableft@skip@float
  \let\tabmid@skip@default\tabmid@skip
  \let\tabmid@skip\tabmid@skip@float
  \let\tabright@skip@default\tabright@skip
  \let\tabright@skip\tabright@skip@float
  \let\array@row@pre@default\array@row@pre
  \let\array@row@pre\array@row@pre@float
  \let\array@row@pst@default\array@row@pst
  \let\array@row@pst\array@row@pst@float
  \appdef\array@row@rst{%
    \let\array@row@pre\array@row@pre@default
    \let\array@row@pst\array@row@pst@default
    \let\tableft@skip\tableft@skip@default
    \let\tabmid@skip\tabmid@skip@default
    \let\tabright@skip\tabright@skip@default
    \appdef\tabular@hook{\let\@halignto\@empty}%
  }%
}{%
}%
\makeatother
\documentclass[fleqn]{iopart}

\usepackage[T1]{fontenc}
\expandafter\let\csname equation*\endcsname\relax
\expandafter\let\csname endequation*\endcsname\relax
\usepackage{amsmath}
\usepackage{amsfonts,amsbsy}
\usepackage{txfonts}
\usepackage{isomath}
\makeatletter
\renewcommand*{\vec}[1]{\ifcat a#1\mathbfit{#1}\else\boldsymbol{#1}\fi}
\makeatother
\usepackage{braket}
\usepackage{microtype}
\usepackage{graphicx}
\graphicspath{{figures/}}
\usepackage[breaklinks=true,
pdfauthor={Heiko Bauke, Sven Ahrens, Christoph H. Keitel, and Rainer Grobe},
pdftitle={What is the relativistic spin operator?}]{hyperref}
\usepackage{ifthen}
\usepackage{textcomp}
\usepackage{subscript}
\usepackage{tabularx}

\ifdefined\Preprint
\else
\fi

\newcommand*{\op}[1]{\hat{#1}}
\newcommand*{\opHo}{\op{H}_0}
\newcommand*{\opHC}{\op{H}_C}
\newcommand*{\opp}{\op{\vec p}}
\newcommand*{\oppo}{\op{p}_0}
\newcommand*{\opsigma}[1][]{\ifthenelse{\equal{#1}{}}{\op{\vec{\sigma}}}{\op{\sigma}_{#1}}}
\newcommand*{\opSigma}[1][]{\ifthenelse{\equal{#1}{}}{\op{\vec{\Sigma}}}{\op{\Sigma}_{#1}}}
\newcommand*{\opS}[1][]{\ifthenelse{\equal{#1}{}}{\op{\vec{S}}}{\op{S}_{#1}}}
\newcommand*{\opL}[1][]{\ifthenelse{\equal{#1}{}}{\op{\vec{L}}}{\op{L}_{#1}}}
\newcommand*{\opJ}[1][]{\ifthenelse{\equal{#1}{}}{\op{\vec{J}}}{\op{J}_{#1}}}
\newcommand*{\opSP}[1][]{\ifthenelse{\equal{#1}{}}{\op{\vec{S}}_\mathrm{P}}{\op{S}_{\mathrm{P},#1}}}
\newcommand*{\opSFW}[1][]{\ifthenelse{\equal{#1}{}}{\op{\vec{S}}_\mathrm{FW}}{\op{S}_{\mathrm{FW},#1}}}
\newcommand*{\opSCz}[1][]{\ifthenelse{\equal{#1}{}}{\op{\vec{S}}_\mathrm{Cz}}{\op{S}_{\mathrm{Cz},#1}}}
\newcommand*{\opSF}[1][]{\ifthenelse{\equal{#1}{}}{\op{\vec{S}}_\mathrm{F}}{\op{S}_{\mathrm{F},#1}}}
\newcommand*{\opSCh}[1][]{\ifthenelse{\equal{#1}{}}{\op{\vec{S}}_\mathrm{Ch}}{\op{S}_{\mathrm{Ch},#1}}}
\newcommand*{\opSPr}[1][]{\ifthenelse{\equal{#1}{}}{\op{\vec{S}}_\mathrm{Pr}}{\op{S}_{\mathrm{Pr},#1}}}
\newcommand*{\opSPnr}[1][]{\ifthenelse{\equal{#1}{}}{\op{\vec{S}}_\mathrm{P,nr}}{\op{S}_{\mathrm{P,nr},#1}}}
\newcommand*{\opSGR}[1][]{\ifthenelse{\equal{#1}{}}{\op{\vec{S}}_\mathrm{GR}}{\op{S}_{\mathrm{GR},#1}}}
\newcommand*{\opSFG}[1][]{\ifthenelse{\equal{#1}{}}{\op{\vec{S}}_\mathrm{FG}}{\op{S}_{\mathrm{FG},#1}}}

\newcommand*{\I}{\mathrm{i}}

\providecommand*{\e}{\mathrm{e}}
\newcommand*{\commut}[2]{%
  \mathchoice{\left[#1,#2\right]}{[#1,#2]}{[#1,#2]}{[#1,#2]}%
}
\newcommand*{\trans}[1]{#1^{\mathsf{T}}}%

\DeclareMathOperator{\RE}{Re}

\ifdefined\ruledtabular\else\newenvironment{ruledtabular}{\footnotesize}{}\fi

\begin{document}

\title{What is the relativistic spin operator?}

\ifdefined\Preprint

\author{Heiko Bauke}
\email{heiko.bauke@mpi-hd.mpg.de}
\affiliation{Max-Planck-Institut f{\"u}r Kernphysik, Saupfercheckweg~1,
  69117~Heidelberg, Germany} 
\author{Sven Ahrens}
\affiliation{Max-Planck-Institut f{\"u}r Kernphysik, Saupfercheckweg~1,
  69117~Heidelberg, Germany} 
\author{Christoph H. Keitel}
\affiliation{Max-Planck-Institut f{\"u}r Kernphysik, Saupfercheckweg~1,
  69117~Heidelberg, Germany} 
\author{Rainer Grobe}
\affiliation{Max-Planck-Institut f{\"u}r Kernphysik, Saupfercheckweg~1,
  69117~Heidelberg, Germany} 
\affiliation{{Intense Laser Physics Theory Unit and Department of
    Physics, Illinois State University, Normal, IL 61790-4560 USA}} 

\else

\author{Heiko Bauke$^1$,
  Sven Ahrens$^1$,
  Christoph H Keitel$^1$ and
  Rainer Grobe$^{1, 2}$}
\address{$^1$ Max-Planck-Institut f{\"u}r Kernphysik, Saupfercheckweg~1,
  69117~Heidelberg, Germany}
\address{$^2$ Intense Laser Physics Theory Unit and Department of
  Physics, Illinois State University, Normal, IL 61790-4560 USA}
\ead{heiko.bauke@mpi-hd.mpg.de}

\fi

\date{\today}

\begin{abstract}
  Although the spin is regarded as a fundamental property of the
  electron, there is no universally accepted spin operator within the
  framework of relativistic quantum mechanics.  We investigate the
  properties of different proposals for a relativistic spin operator.
  It is shown that most candidates are lacking essential features of
  proper angular momentum operators, leading to spurious Zitterbewegung
  (quivering motion) or violating the angular momentum algebra.  Only
  the Foldy-Wouthuysen operator and the Pryce operator qualify as
  proper relativistic spin operators.  We demonstrate that ground
  states of highly charged hydrogen-like ions can be utilized to
  identify a legitimate relativistic spin operator experimentally.

  \vspace{\baselineskip}%
  \noindent Keywords: spin, relativistic quantum mechanics,
  hydrogen-like ions
  \vspace{-2.5\baselineskip}%
\end{abstract}

\pacs{03.65.Pm, 31.15.aj, 31.30.J-}

\ifdefined\Preprint
\maketitle
\fi

\section{Introduction}
\label{sec:intro}%

Quantum mechanics forms the universally accepted theory for the
description of physical processes on the atomic scale.  It has been
validated by countless experiments and it is used in many technical
applications.  However, quantum mechanics presents physicists with
some conceptual difficulties even today.  In particular, the concept
of spin is related to such difficulties and myths
\cite{Nikolic:2007:Myths_Facts,Giulini:2008:spin}.  Although there is
consensus that elementary particles have some quantum mechanical
property that is called spin, the understanding of the physical nature
of the spin is still incomplete \cite{Morrison:2007:Spin}.

Historically, the concept of spin was introduced in order to explain
some experimental findings such as the emission spectra of alkali metals
and the Stern-Gerlach experiment.  A direct measuring of the spin (or
more precisely the electron's magnetic moment), however, was missing
until the pioneering work by Dehmelt \cite{Dehmelt:1990:Structure}.
Nevertheless, spin measurement experiments
\cite{Hanneke:etal:2008:New_Measurement,Neumann:etal:2010:Single-Shot_Readout,Buckley:2010:Spin-Light_Coherence,Close:etal:2011:Spin_State_Amplification,Sturm:etal:2011:g_factor,DiSciacca:2012:Direct_Measurement}
still require sophisticated methods. Pauli and Bohr even claimed that
the spin of free electrons was impossible to measure for fundamental
reasons \cite{Mott:1929:Scattering}.  Recent renewed interest in
fundamental aspects of the spin arose, for example, from the growing
field of (relativistic) quantum information
\cite{Peres:Terno:2004:QI_RT,Kim:etal:2005:Bell,Wiesendanger:2009:Spin_mapping,Friis:etal:2010,Petersson:2010:Circuit_qed,Saldanha:Vedral:2012:Wigner},
quantum spintronics \cite{Awschalom:etal:2013:Quantum_Spintronics}, spin
effects in graphene
\cite{Abanin:2011:Giant_Spin-Hall_Effect,Mecklenburg:Regan:2011:Spin,Guttinger:2010:Spin_States}
and in light-matter interaction at relativistic intensities
\cite{DiPiazza:etal:2012:high-intensity,Ahrens:etal:2012:spin_dynamics,Ahrens:etal:2014:Electron-spin_dynamics_induced_by_photon_spins}.

According to the formalism of quantum mechanics, each measurable
quantity is represented by a Hermitian operator.  Taking the
experiments that aim to measure bare electron spins seriously, we have
to ask the question: What is the correct (relativistic) spin operator.
Although the spin is regarded as a fundamental property of the
electron, a universally accepted spin operator for the Dirac theory is
still missing.  The pivotal question we try to tackle is: Which
mathematical operator corresponds to an experimental spin measurement?
This question may be answered by comparing experimental results with
theoretical predictions originating from different spin operators and
testing which operator is compatible with experimental data.

A relativistic spin operator may be introduced by splitting the
undisputed total angular momentum operator $\opJ$ into an external
part $\opL$ and an internal part $\opS$, commonly referred to as the
orbital angular momentum and the spin, viz. $\opJ=\opL+\opS$.  The
question for the right splitting of the total angular momentum into an
orbital part and a spin part is closely related to the quest for the
right relativistic position operator
\cite{Newton:Wigner:1949:Localized_States,Jordan:Mukunda:1963:Lorentz-Covariant_Position_Operators,OConnell:Wigner:1978:Position_operators}.
This becomes evident by writing $\opL=\op{\vec{r}}\times\op{\vec{p}}$
with the position operator $\op{\vec{r}}$ and the kinematic momentum
operator $\op{\vec{p}}$, which is in atomic units as used in this
paper $\op{\vec{p}}=-\I\vec\nabla$.  Thus, different definitions of
the spin operator $\opS$ induce different relativistic position
operators $\op{\vec r}$.

Introducing the position vector $\vec r$ and the operator $\op{\vec
  \Sigma}=\trans{(\op\Sigma_1, \op\Sigma_2, \op\Sigma_3)}$ via
\begin{equation}
  \op\Sigma_i = -\I\alpha_j \alpha_k
\end{equation}
with $(i,j,k)$ being a cyclic permutation of $(1,2,3)$ and the matrices
$\trans{(\alpha_1, \alpha_2, \alpha_3)}=\vec\alpha$ obeying the algebra
\begin{equation}
  \label{eq:Dirac_algebra}
  \alpha_i^2 = 1\,,\qquad
  \alpha_i\alpha_k + \alpha_k\alpha_i = 2\delta_{i,k}\,,
\end{equation}
the operator of the relativistic total angular momentum is given by
$\opJ=\vec r\times\opp+\op{\vec{\Sigma}}/2$.  Thus, the most obvious
way of splitting $\opJ$ is to define the orbital angular momentum
operator $\op{\vec L}_\mathrm{P}=\vec r\times\opp$ and the spin
operator $\opSP=\opSigma/2$, which is a direct generalization of the
orbital angular momentum operator and the spin operator of the
nonrelativistic Pauli theory.  This naive splitting, however, suffers
from several problems, e.\,g., $\op{\vec L}_\mathrm{P}$ and $\opSP$ do
not commute with the free Dirac Hamiltonian nor with the Dirac
Hamiltonian for central potentials.  Thus, in contrast to classical
and nonrelativistic quantum theory the angular momenta $\op{\vec
  L}_\mathrm{P}$ and $\opSP$ are not conserved.  This has
consequences, e.\,g., for the labeling of the eigenstates of the
hydrogen atom. In nonrelativistic theory, bound hydrogen states may be
constructed as simultaneous eigenstates of the Pauli-Coulomb
Hamiltonian, the squared orbital angular momentum, the
\mbox{$z$-components} of the orbital angular momentum and the spin. In
the Dirac theory, however, the squared total angular momentum
$\opJ^2$, the total angular momentum in \mbox{$z$-direction}
$\opJ[3]$, and the so-called spin-orbit operator $\op K$ (or the
parity) are utilized
\cite{Bethe:Salpeter:2008:Atoms,Thaller:2005:vis_QM_II}. In
particular, it is not possible to construct simultaneous eigenstates
of the Dirac-Coulomb Hamiltonian and some component of $\opSP$.

\section{Relativistic spin operators}
\label{sec:operators}

\renewcommand{\tabularxcolumn}[1]{m{#1}}

\begin{table*}
  \centering
  \caption{Definitions and commutation properties of various
    relativistic spin operators.} 
  \begin{ruledtabular}
  \begin{tabularx}{\textwidth}{@{}Xllll@{}}
    \br
    &  & $\commut{\opHo}{\opS}$ &
      $\commut{\opS[i]}{\opS[j]}$ &
      Eigenvalues \\
      Operator name & Definition & $=0$? &
      $=\I\varepsilon_{i,j,k}\opS[k]$? &
      $=\pm1/2$? \\
      \mr
      Pauli 
      \cite{Hill:Landshoff:1938,Dirac:1971:Positive-Energy,Ohanian:1986:spin} & 
      $\opSP = \dfrac 12 \opSigma$ &
      no & yes & yes \\[1.75ex]
      Foldy-Wouthuysen 
      \cite{Foldy:Wouthuysen:1950:Non-Relativistic_Limit,deVries:1970:FW_Transformation,Costella:McKellar:1995:Foldy-Wouthuysen,Schweber:2005:qft,Caban:etal:2013:A_spin_observable}
      & 
      $ \opSFW = \dfrac{1}{2}\opSigma + 
      \dfrac{\I\beta}{2\oppo}\opp\times\vec\alpha -
      \dfrac{\opp\times(\opSigma\times\opp)}{2\oppo(\oppo+m_0c)}$ & 
      yes & yes & yes \\[1.25ex]
      Czachor \cite{Czachor:1997:relativistic_EPR} & 
      $\opSCz 
      = \dfrac{m_0^2c^2}{2\oppo^2}\opSigma + 
      \dfrac{\I m_0c\beta}{2\oppo^2}\opp\times\vec\alpha +
      \dfrac{\opp\cdot\opSigma}{2\oppo^2}\opp$ & 
      yes & no & no \\[1.75ex]
      Frenkel 
      \cite{Hilgevoord:Wouthuysen:1963:Dirac_spin,Wightman:1960,Bargmann:etal:1935:Precession} & 
      $\opSF = \dfrac{1}{2}\opSigma +
      \dfrac{\I\beta}{2m_0c}\opp\times\vec\alpha$ & 
      yes & no & no \\[1.75ex]
      Chakrabarti 
      \cite{Chakrabarti:1963:Canonical_Form,Gursey:1965:Equivalent,Gursey:1965,Gursey:1965:Equivalent_formulations,Kirsch:Ryder:Hehl:2001:Gordon_decompositions,Ryder:1999:Relativistic_Spin,Choi:2013:Relativistic_spin_operator} & 
      $ \opSCh = 
      \dfrac{1}{2}\opSigma + 
      \dfrac{\I}{2m_0c}\vec\alpha\times\opp +
      \dfrac{\opp\times(\opSigma\times\opp)}{2m_0c(m_0c+\oppo)}$ &
      no & yes & yes \\[1.75ex]
      Pryce 
      \cite{Ryder:1999:Relativistic_Spin,Pryce:1935:Commuting_Co-Ordinates,Pryce:1948:Mass-Centre,Macfarlane:1963:Relativistic_Two_Particle,Berg:1965:Position_Spin} & 
      $ \opSPr = 
      \dfrac{1}{2}\beta\opSigma + 
      \dfrac{1}{2}\opSigma\cdot\opp(1-\beta)\dfrac{\opp}{\opp^2}$ & 
      yes & yes & yes \\[1.75ex]
      Fradkin-Good
      \cite{Kirsch:Ryder:Hehl:2001:Gordon_decompositions,Fradkin:Good:1961:Polarization_Operators} &
      $ \opSFG = 
      \dfrac{1}{2}\beta\opSigma + 
      \dfrac{1}{2}\opSigma\cdot\opp\left(\dfrac{\opHo}{c\oppo}-\beta\right)\dfrac{\opp}{\opp^2}$ & 
      yes & no & yes \\[1.75ex]
      \br
    \end{tabularx}
  \end{ruledtabular}
  \label{tab:spin_operators}
\end{table*}

To overcome conceptual problems with the naive splitting of $\opJ$ into
$\op{\vec L}_\mathrm{P}$ and $\opSP$, several alternatives for a
relativistic spin operator have been proposed. However, there is no
single commonly accepted relativistic spin operator, leading to the
unsatisfactory situation that the relativistic spin operator is not
unambiguously defined.  We will investigate the properties of different
popular definitions of the spin operator which result from different
splittings of $\opJ$ with the aim to find means that allow to identify
the legitimate relativistic spin operator by experimental methods.

Table~\ref{tab:spin_operators} summarizes various proposals for a
relativistic spin operator $\opS$.  These operators are often motivated
by abstract group theoretical considerations rather than by experimental
evidence.  For example, Wigner showed in his seminal work
\cite{Wigner:1939:Unitary_Representations,Bargmann:Wigner:1948:Group_Theoretical_Discussion,Wigner:1957:Relativistic_Invariance}
that the spin degree of freedom can be associated with irreducible
representations of the sub-group of the inhomogeneous Lorentz group that
leaves the four-momentum invariant.  We will denote individual
components of $\opS$ by $\opS[i]$ with index $i\in\{1,2,3\}$.  The spin
operators are defined in terms of the particle's rest mass $m_0$, the
speed of light $c$, the matrix $\beta$ such that
\begin{equation}
  \beta^2 = 1 \,,\qquad \alpha_i\beta + \beta\alpha_i = 0\,,
\end{equation}
the free particle Dirac Hamiltonian
\begin{align}
  \op H_0 & = c\vec\alpha\cdot\opp + m_0c^2\beta \,,\\
  \intertext{and the operator}
  \oppo & = (m_0^2c^2 + \opp^2)^{1/2} \,.
\end{align}
In the nonrelativistic limit, i.\,e., when the plane wave expansion of
a wave packet has only components with momenta which are small
compared to $m_0c$, expectation values for all operators in
Tab.~\ref{tab:spin_operators} converge to the same value.  Note that
the nomenclature in Tab.~\ref{tab:spin_operators} is not universally
adopted in the literature and other authors may utilize different
operator names.  Furthermore, the spin operators can be formulated by
various different but algebraically equivalent expressions.  For
example, the so-called G{\"u}rsey-Ryder operator in
\cite{Kirsch:Ryder:Hehl:2001:Gordon_decompositions,Ryder:1999:Relativistic_Spin}
is equivalent to the Chakrabarti operator of
Tab.~\ref{tab:spin_operators}.

One may conclude that an operator can not be considered as a
relativistic spin operator if it does not inherit the key properties of
the nonrelativistic Pauli spin operator.  In particular, we demand from
a proper relativistic spin operator the following features:
\begin{enumerate}
\item It is required to commute with the free Dirac Hamiltonian.
\item A spin operator must feature the two eigenvalues $\pm1/2$ and it
  has to obey the angular momentum algebra
  \begin{equation}
    \label{eq:commut_S}
    \textstyle\commut{\opS[i]}{\opS[j]}=\I\varepsilon_{i,j,k}\opS[k]
  \end{equation}
  with $\varepsilon_{i,j,k}$ denoting the Levi-Civita symbol.
\end{enumerate}
The first property is required to ensure that the relativistic spin
operator is a constant of motion if forces are absent, such that spurious
Zitterbewegung of the spin is prevented.  The second requirement is
commonly regarded as \emph{the} fundamental property of angular momentum
operators of spin-half particles
\cite{Sakurai:Napolitano:2010:Modern_QM}.  The physical quantity that is
represented by the operator $\opS$ should not depend on the orientation
of the chosen coordinate system.  This can be ensured by fulfilling 
\cite{Sakurai:Napolitano:2010:Modern_QM}
\begin{equation}
  \label{eq:commut_J_S}
  \textstyle\commut{\op{J}_i}{\opS[j]} =
  \I\varepsilon_{i,j,k}\opS[k]\,.
\end{equation}
The angular momentum algebra (\ref{eq:commut_S}) and the relation
(\ref{eq:commut_J_S}) determine the properties of the spin and the
orbital angular momentum as well as the relationship between them.  As
a consequence of (\ref{eq:commut_J_S}), the orbital angular momentum
$\opL=\opJ-\opS$ that is induced by a particular choice of the spin
obeys $\commut{\opJ[i]}{\opL[j]} = \I\varepsilon_{i,j,k}\opL[k]$.
Thus, $\opL$ is a physical vector operator, too.  As $\opL$ represents
an angular momentum operator, it must obey the angular momentum
algebra.  Furthermore, we may say that the total angular momentum
$\opJ$ is split into an internal part $\opS$ and an external part
$\opL$ only if internal and external angular momenta can be measured
independently, i.\,e., $\opS$ and $\opL$ commute.  Both conditions are
fulfilled if, and only if, the spin operator $\opS$ satisfies the
angular momentum algebra (\ref{eq:commut_S}) because the commutator
relations
\begin{align}
  \textstyle\commut{\opL[i]}{\opL[j]} & = \I\varepsilon_{i,j,k}\opL[k] +
  \textstyle\commut{\opS[i]}{\opS[j]} - \I\varepsilon_{i,j,k}\opS[k] \,,\\
  \textstyle\commut{\opL[i]}{\opS[j]} & = \I\varepsilon_{i,j,k}\opS[k] -
  \textstyle\commut{\opS[i]}{\opS[j]}
\end{align}
follow from (\ref{eq:commut_J_S}).  All spin operators in
Tab.~\ref{tab:spin_operators} fulfill (\ref{eq:commut_J_S}).  The
Czachor spin operator $\opSCz$, the Frenkel spin operator $\opSF$, and
the Fradkin-Good operator $\opSFG$, however, are disqualified as
relativistic spin operators by violating the angular momentum algebra
(\ref{eq:commut_S}).  Furthermore, the Pauli spin operator $\opSP$ and
the Chakrabarti spin operator $\opSCh$ do not commute with the free
Dirac Hamiltonian, ruling them out as meaningful relativistic spin
operators.  According to our criteria, only the Foldy-Wouthuysen spin
operator $\opSFW$ and the Pryce spin operator $\opSPr$ remain as
possible relativistic spin operators.

\section{Electron spin of hydrogen-like ions}
\label{sec:hydrogen}

The question of which of the proposed relativistic spin operators (if
any) in Tab.~\ref{tab:spin_operators} provides the correct
mathematical description of spin can be answered definitely only by
comparing theoretical predictions with experimental results.  For this
purpose one needs a physical setup that shows strong relativistic
effects and is as simple as possible. Such a setup is provided by the
bound eigenstates of highly charged hydrogen-like ions, i.\,e., atomic
systems with an atomic core of $Z$ protons and a single electronic
charge.  These ions can be produced at storage rings
\cite{Stohlker:etal:1993:Ground-state_Lamb_shift} or by utilizing
electron beam ion traps
\cite{Robbins:etal:2006:Polarization_measurements,Kluge:etal:2008:HITRAP}
up to $Z=92$ (hydrogen-like uranium).  The degenerate bound
eigenstates of the corresponding Coulomb-Dirac Hamiltonian
\begin{equation}
  \opHC = \opHo - \frac{Z}{|\vec r|} 
\end{equation}
are commonly expressed as simultaneous eigenstates
$\psi_{n,j,m,\kappa}$ of $\opHC$, $\opJ^2$, $\opJ[3]$, and the
so-called spin-orbit operator $\op{K}=\mbox{$\beta\{\opSigma\cdot[\vec
  r\times(-\I\vec\nabla)+1)]\}$}$ fulfilling the eigenequations
\cite{Bethe:Salpeter:2008:Atoms,Thaller:2005:vis_QM_II}
\begin{align}
  \opHC \psi_{n,j,m,\kappa} & = \mathcal{E}(n, j)\psi_{n,j,m,\kappa}
  &
  n & = 1,2,\dots\,,\\
  \opJ^2 \psi_{n,j,m,\kappa} & = j(j+1)\psi_{n,j,m,\kappa} &
  j & = \tfrac{1}{2},\tfrac{3}{2},\dots,n-\tfrac{1}{2} \,,\\
  \opJ[3] \psi_{n,j,m,\kappa} & = m\psi_{n,j,m,\kappa} &
  m & = -j,(j-1),\dots,j \,,\\
  \op{K} \psi_{n,j,m,\kappa} & = \kappa\psi_{n,j,m,\kappa} &
  \kappa & = -j-\tfrac{1}{2},j+\tfrac{1}{2}\,.
\end{align}
The eigenenergies are given with $\alpha_\mathrm{el}$ denoting the fine
structure constant by
\begin{equation}
  \label{eq:energy}
  \mathcal{E}(n, j) = 
  m_0c^2\left[
    1 + \left(
      \dfrac{\alpha_\mathrm{el}^2Z^2}
      {n-j-{1}/{2}+\sqrt{\textstyle(j-{1}/{2})^2-\alpha^2_\mathrm{el}Z^2}}
    \right)
  \right]^{-{1}/{2}}\,.
\end{equation}

In order to establish a close correspondence between the nonrelativistic
Schr{\"o}dinger-Pauli theory and the relativistic Dirac theory, one may
desire to find a splitting of $\opJ$ into a sum $\opJ=\opL+\opS$ of
commuting operators such that both $\opL$ and $\opS$
\begin{enumerate}
\item fulfill the angular momentum algebra, and 
\item form a complete set of commuting operators that contains $\opHC$ as
  well as $\opS[3]$ and/or $\opL[3]$.
\end{enumerate}
The latter property would ensure that \emph{all} hydrogenic energy
eigenstates are spin eigenstates and/or orbital angular momentum
eigenstates, too.  Such hypothetical eigenstates would be
superpositions of $\psi_{n,j,m,\kappa}$ of the same
energy. Consequently, these superpositions are eigenstates of
$\opJ^2$, too, because the energy (\ref{eq:energy}) depends on the
principal quantum number $n$ as well as the quantum number $j$.  Thus,
any complete set of commuting operators for specifying hydrogenic
quantum states necessarily includes $\opJ^2$. As a consequence of the
postulated angular momentum algebra for $\opL$ and $\opS$, the
operator $\opJ^2$ commutes with $\opL^2$ as well as with $\opS^2$, but
with neither $\opS[3]$ nor $\opL[3]$
\cite{Cohen-Tannoudji:etal:1997:QM_VII} excluding $\opS[3]$ and
$\opL[3]$ from any complete set of commuting operators for specifying
relativistic hydrogenic eigenstates.  In conclusion, hydrogenic energy
eigenstates are generally not eigenstates of any spin operator that
fulfills the angular momentum algebra.

In momentum space, the relativistic spin operators introduced in
Tab.~\ref{tab:spin_operators} are simple matrices. Thus, by employing
the momentum space representation of $\psi_{n,j,m,\kappa}$, spin
expectation values of the degenerate hydrogenic ground states
$\psi_\uparrow=\psi_{1,1/2,1/2,1}$ and
$\psi_\downarrow=\psi_{1,1/2,-1/2,-1}$ can be evaluated
\cite{Rubinowicz:1948:Coulomb_Momentum}.  For simplicity, we measure
spin along the \mbox{$z$-direction} for the reminder of this section.
The spin expectation value of a general superposition %
$\psi=\cos({\eta}/{2})\psi_{\uparrow} +
\sin({\eta}/{2})\e^{\I\zeta}\psi_{\downarrow}$ of the hydrogenic
ground states $\psi_\uparrow$ and $\psi_\downarrow$ is given by
\begin{multline}
  \label{eq:mean_S3}
  \Braket{ \psi | \opS[3] | \psi} 
  = \cos^2\frac{\eta}{2}\braket{\psi_{\uparrow} | \opS[3] |
    \psi_{\uparrow} } +
  \sin^2\frac{\eta}{2}\braket{\psi_{\downarrow} | \opS[3] | \psi_{\downarrow} } \\
  + 2 \cos\frac{\eta}{2}\sin\frac{\eta}{2}\cos\zeta
  \RE\braket{\psi_{\uparrow} | \opS[3] | \psi_{\downarrow}
  }\,.
\end{multline}
For all spin operators introduced in Tab.~\ref{tab:spin_operators}, the
mixing term $\RE\braket{\psi_{\uparrow} | \opS[3] | \psi_{\downarrow}}$
vanishes and, furthermore, %
$\braket{\psi_{\uparrow} | \opS[3] | \psi_{\uparrow} }=
-\braket{\psi_{\downarrow} | \opS[3] | \psi_{\downarrow} }>0$.  Thus,
the expectation value (\ref{eq:mean_S3}) is maximal for $\eta=0$ and
minimal for $\eta=\pi$ and the inequality
\begin{equation}
  \label{eq:bounds}
  \braket{\psi_{\downarrow} | \opS[3] | \psi_{\downarrow}} \le
  \braket{\psi | \opS[3] | \psi} \le
  \braket{\psi_{\uparrow} | \opS[3] | \psi_{\uparrow}}  
\end{equation}
holds for all hydrogenic ground states $\psi$.

\begin{figure}
  \centering
  \ifdefined\Preprint
  \includegraphics[width=\linewidth]{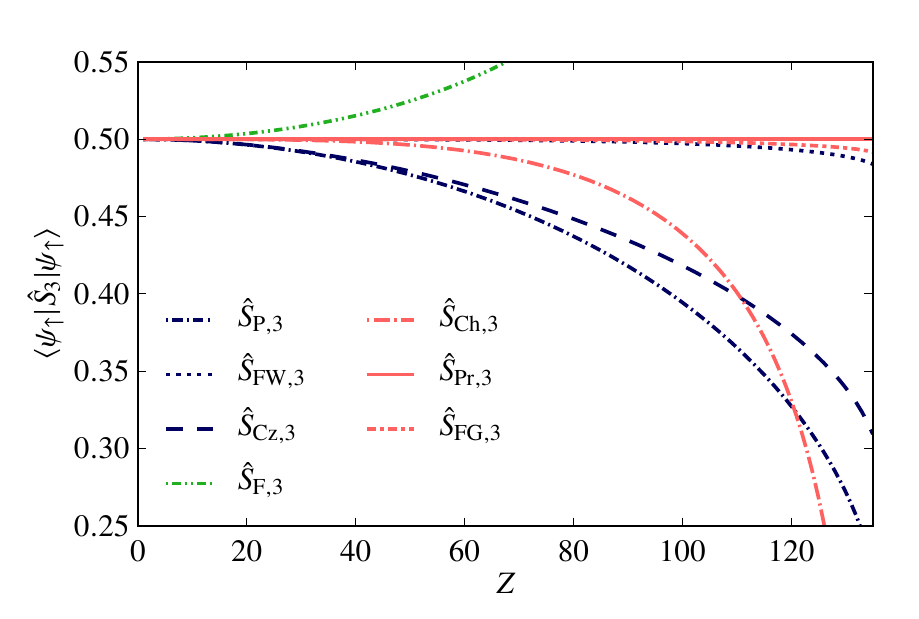}%
  \else
  \includegraphics[scale=0.575]{hydrogen_spin}%
  \fi
  \caption{Spin expectation values
    $\braket{\psi_{\uparrow}|\opS[3]|\psi_{\uparrow}}$ of various
    relativistic spin operators for the hydrogenic ground state
    $\psi_\uparrow$ as a function of the atomic number $Z$ measured in
    \mbox{$z$-direction}. Spin expectation values for $\psi_\downarrow$
    follow by symmetry via
    $\braket{\psi_{\uparrow}|\opS[3]|\psi_{\uparrow}}=
    -\braket{\psi_{\downarrow}|\opS[3]|\psi_{\downarrow}}$.}
  \label{fig:hydrogen_spin}
\end{figure}

For every proposed spin operator in Tab.~\ref{tab:spin_operators} we get
different values for the upper and lower bounds in (\ref{eq:bounds}).
The spin expectation values
$\braket{\psi_{\uparrow}|\opS[3]|\psi_{\uparrow}}$ and implicitly
$\braket{\psi_{\downarrow}|\opS[3]|\psi_{\downarrow}}$ for the operators
of Tab.~\ref{tab:spin_operators} are displayed as a function of the
atomic number $Z$ in Fig.~\ref{fig:hydrogen_spin}.  None of the spin
operators in Tab.~\ref{tab:spin_operators} commutes with $\opHC$. Thus,
the expectation values $\braket{\psi_{\uparrow}|\opS[3]|\psi_{\uparrow}
}$ and $\braket{\psi_{\downarrow}|\opS[3]|\psi_{\downarrow}}$ generally
do not equal one of the eigenvalues of $\opS[3]$.  For small atomic
numbers ($Z<20$), all spin operators yield about $\pm1/2$.  For larger
$Z$, however, expectation values differ significantly from each other.
In particular, spin expectation values differ from $\pm1/2$ even for
spin operators with eigenvalues $\pm1/2$.  This means, it is possible to
discriminate between different relativistic spin operator candidates.
The spin expectation value's magnitude decreases with growing $Z$ when
the Pauli, the Fouldy-Wouthuysen, the Czachor, the Chakrabarti, or the
Fradkin-Good spin operator is applied. The Frenkel spin operator yields
spin expectation values with modulus exceeding $1/2$ which is due to the
violation of the angular momentum algebra.  Only the Pryce operator
yields spin expectation of $\pm1/2$ for all values of $Z$.  In fact,
calculations show that all hydrogenic states $\psi_{n,j,m,\kappa}$ with
$m=\pm j$ are eigenstates of the Pryce spin operator but not those
states with $m\ne\pm j$.

\section{An experimental test for relativistic spin operators}
\label{sec:test}

Theoretical considerations led to several proposals for a relativistic
spin operator as illustrated in Tab.~\ref{tab:spin_operators}.  The
identification of the correct relativistic spin operator, however,
demands an experimental test.  The inequality (\ref{eq:bounds}) may
serve as a basis for such an experimental test.  More precisely, the
inequality (\ref{eq:bounds}) allows falsification of the hypothesis
that the spin measurement procedure is an experimental realization of
some operator $\opS$, where $\opS$ is one of the operators in
Tab.~\ref{tab:spin_operators}.  In this test the electron of a highly
charged hydrogen-like ion is prepared in its ground state
$\psi_\uparrow$ first, e.\,g., by exposing the ion to a strong
magnetic field in \mbox{$z$-direction} and turning it off
adiabatically.  (Preparing a superposition of \mbox{$\psi_\uparrow$}
and \mbox{$\psi_\downarrow$} will reduce the sensitivity of the
experimental test.) Afterwards, the spin will be measured along
\mbox{$z$-direction}, e.\,g., by a Stern-Gerlach-like experiment,
yielding the experimental expection value $s$.  Comparing this
experimental value to each of the seven bounds shown in
Fig.~\ref{fig:hydrogen_spin} will allow exclusion of some of the
proposed spin operators.  The hypothesis that the spin measurement
procedure is an experimental realization of the operator $\opS$ is
compatible with the experimental result $s$ if, and only if, the
inequality
\begin{math}
  \braket{\psi_{\downarrow} | \opS[3] | \psi_{\downarrow}} \le
  s\le
  \braket{\psi_{\uparrow} | \opS[3] | \psi_{\uparrow}}  
\end{math}
is fulfilled.  Otherwise, this operator is excluded as a relativistic
spin operator by experimental evidence.  In particular, realizing full
spin-polarization, i.\,e., \mbox{$s=\pm1/2$}, eliminates all operators
in Tab.~\ref{tab:spin_operators} except the Pryce operator.

\section{Conclusions}
\label{sec:conclusion}

We investigated the properties of various proposals for a relativistic
spin operator.  Only the Fouldy-Wouthuysen operator and the Pryce
operator fulfill the angular momentum algebra and are constants of
motion in the absence of forces.  While different theoretical
considerations led to different spin operators, the definite
relativistic spin operator has to be justified by experimental evidence.
Energy eigenstates of highly charged hydrogen-like ions, in particular
the ground states, can be utilized to exclude candidates for a
relativistic spin operator experimentally.  The proposed spin operators
predict different maximal degrees of spin polarization.  Only the Pryce
spin operator allows for a complete polarization of spin in the
hydrogenic ground state.

\ack We have enjoyed helpful discussions with Prof.~C.~M{\"u}ller,
S.~Meuren, Prof.~Q.~Su and E.~Yakaboylu.  R.~G.\ acknowledges the warm
hospitality during his sabbatical leave in Heidelberg.  This work was
supported by the NSF.

\ifdefined\Preprint
\else
\section*{References}
\fi

\bibliography{relativistic_hydrogen_spin}

\begin{thebibliography}{62}%
\makeatletter
\providecommand \@ifxundefined [1]{%
 \@ifx{#1\undefined}
}%
\providecommand \@ifnum [1]{%
 \ifnum #1\expandafter \@firstoftwo
 \else \expandafter \@secondoftwo
 \fi
}%
\providecommand \@ifx [1]{%
 \ifx #1\expandafter \@firstoftwo
 \else \expandafter \@secondoftwo
 \fi
}%
\providecommand \natexlab [1]{#1}%
\providecommand \enquote  [1]{``#1''}%
\providecommand \bibnamefont  [1]{#1}%
\providecommand \bibfnamefont [1]{#1}%
\providecommand \citenamefont [1]{#1}%
\providecommand \href@noop [0]{\@secondoftwo}%
\providecommand \href [0]{\begingroup \@sanitize@url \@href}%
\providecommand \@href[1]{\@@startlink{#1}\@@href}%
\providecommand \@@href[1]{\endgroup#1\@@endlink}%
\providecommand \@sanitize@url [0]{\catcode `\\12\catcode `\$12\catcode
  `\&12\catcode `\#12\catcode `\^12\catcode `\_12\catcode `\%12\relax}%
\providecommand \@@startlink[1]{}%
\providecommand \@@endlink[0]{}%
\providecommand \url  [0]{\begingroup\@sanitize@url \@url }%
\providecommand \@url [1]{\endgroup\@href {#1}{\urlprefix }}%
\providecommand \urlprefix  [0]{URL }%
\providecommand \Eprint [0]{\href }%
\providecommand \doibase [0]{http://dx.doi.org/}%
\providecommand \selectlanguage [0]{\@gobble}%
\providecommand \bibinfo  [0]{\@secondoftwo}%
\providecommand \bibfield  [0]{\@secondoftwo}%
\providecommand \translation [1]{[#1]}%
\providecommand \BibitemOpen [0]{}%
\providecommand \bibitemStop [0]{}%
\providecommand \bibitemNoStop [0]{.\EOS\space}%
\providecommand \EOS [0]{\spacefactor3000\relax}%
\providecommand \BibitemShut  [1]{\csname bibitem#1\endcsname}%
\let\auto@bib@innerbib\@empty
\bibitem [{\citenamefont {Nikoli\'c}(2007)}]{Nikolic:2007:Myths_Facts}%
  \BibitemOpen
  \bibfield  {author} {\bibinfo {author} {\bibfnamefont {H.}~\bibnamefont
  {Nikoli\'c}},\ }\href {\doibase 10.1007/s10701-007-9176-y} {\bibfield
  {journal} {\bibinfo  {journal} {Foundations of Physics}\ }\textbf {\bibinfo
  {volume} {37}},\ \bibinfo {pages} {1563} (\bibinfo {year}
  {2007})}\BibitemShut {NoStop}%
\bibitem [{\citenamefont {Giulini}(2008)}]{Giulini:2008:spin}%
  \BibitemOpen
  \bibfield  {author} {\bibinfo {author} {\bibfnamefont {D.}~\bibnamefont
  {Giulini}},\ }\href {\doibase 10.1016/j.shpsb.2008.03.005} {\bibfield
  {journal} {\bibinfo  {journal} {Studies In History and Philosophy of Science
  Part B: Studies In History and Philosophy of Modern Physics}\ }\textbf
  {\bibinfo {volume} {39}},\ \bibinfo {pages} {557} (\bibinfo {year}
  {2008})}\BibitemShut {NoStop}%
\bibitem [{\citenamefont {Morrison}(2007)}]{Morrison:2007:Spin}%
  \BibitemOpen
  \bibfield  {author} {\bibinfo {author} {\bibfnamefont {M.}~\bibnamefont
  {Morrison}},\ }\href {\doibase 10.1016/j.shpsb.2006.10.003} {\bibfield
  {journal} {\bibinfo  {journal} {Studies In History and Philosophy of Science
  Part B: Studies In History and Philosophy of Modern Physics}\ }\textbf
  {\bibinfo {volume} {38}},\ \bibinfo {pages} {529} (\bibinfo {year}
  {2007})}\BibitemShut {NoStop}%
\bibitem [{\citenamefont {Dehmelt}(1990)}]{Dehmelt:1990:Structure}%
  \BibitemOpen
  \bibfield  {author} {\bibinfo {author} {\bibfnamefont {H.}~\bibnamefont
  {Dehmelt}},\ }\href {\doibase 10.1126/science.247.4942.539} {\bibfield
  {journal} {\bibinfo  {journal} {Science}\ }\textbf {\bibinfo {volume}
  {247}},\ \bibinfo {pages} {539} (\bibinfo {year} {1990})}\BibitemShut
  {NoStop}%
\bibitem [{\citenamefont {Hanneke}\ \emph {et~al.}(2008)\citenamefont
  {Hanneke}, \citenamefont {Fogwell},\ and\ \citenamefont
  {Gabrielse}}]{Hanneke:etal:2008:New_Measurement}%
  \BibitemOpen
  \bibfield  {author} {\bibinfo {author} {\bibfnamefont {D.}~\bibnamefont
  {Hanneke}}, \bibinfo {author} {\bibfnamefont {S.}~\bibnamefont {Fogwell}}, \
  and\ \bibinfo {author} {\bibfnamefont {G.}~\bibnamefont {Gabrielse}},\ }\href
  {\doibase 10.1103/PhysRevLett.100.120801} {\bibfield  {journal} {\bibinfo
  {journal} {Phys. Rev. Lett.}\ }\textbf {\bibinfo {volume} {100}},\ \bibinfo
  {pages} {120801} (\bibinfo {year} {2008})}\BibitemShut {NoStop}%
\bibitem [{\citenamefont {Neumann}\ \emph {et~al.}(2010)\citenamefont
  {Neumann}, \citenamefont {Beck}, \citenamefont {Steiner}, \citenamefont
  {Rempp}, \citenamefont {Fedder}, \citenamefont {Hemmer}, \citenamefont
  {Wrachtrup},\ and\ \citenamefont
  {Jelezko}}]{Neumann:etal:2010:Single-Shot_Readout}%
  \BibitemOpen
  \bibfield  {author} {\bibinfo {author} {\bibfnamefont {P.}~\bibnamefont
  {Neumann}}, \bibinfo {author} {\bibfnamefont {J.}~\bibnamefont {Beck}},
  \bibinfo {author} {\bibfnamefont {M.}~\bibnamefont {Steiner}}, \bibinfo
  {author} {\bibfnamefont {F.}~\bibnamefont {Rempp}}, \bibinfo {author}
  {\bibfnamefont {H.}~\bibnamefont {Fedder}}, \bibinfo {author} {\bibfnamefont
  {P.~R.}\ \bibnamefont {Hemmer}}, \bibinfo {author} {\bibfnamefont
  {J.}~\bibnamefont {Wrachtrup}}, \ and\ \bibinfo {author} {\bibfnamefont
  {F.}~\bibnamefont {Jelezko}},\ }\href {\doibase 10.1126/science.1189075}
  {\bibfield  {journal} {\bibinfo  {journal} {Science}\ }\textbf {\bibinfo
  {volume} {329}},\ \bibinfo {pages} {542} (\bibinfo {year}
  {2010})}\BibitemShut {NoStop}%
\bibitem [{\citenamefont {Buckley}\ \emph {et~al.}(2010)\citenamefont
  {Buckley}, \citenamefont {Fuchs}, \citenamefont {Bassett},\ and\
  \citenamefont {Awschalom}}]{Buckley:2010:Spin-Light_Coherence}%
  \BibitemOpen
  \bibfield  {author} {\bibinfo {author} {\bibfnamefont {B.~B.}\ \bibnamefont
  {Buckley}}, \bibinfo {author} {\bibfnamefont {G.~D.}\ \bibnamefont {Fuchs}},
  \bibinfo {author} {\bibfnamefont {L.~C.}\ \bibnamefont {Bassett}}, \ and\
  \bibinfo {author} {\bibfnamefont {D.~D.}\ \bibnamefont {Awschalom}},\ }\href
  {\doibase 10.1126/science.1196436} {\bibfield  {journal} {\bibinfo  {journal}
  {Science}\ }\textbf {\bibinfo {volume} {330}},\ \bibinfo {pages} {1212}
  (\bibinfo {year} {2010})}\BibitemShut {NoStop}%
\bibitem [{\citenamefont {Close}\ \emph {et~al.}(2011)\citenamefont {Close},
  \citenamefont {Fadugba}, \citenamefont {Benjamin}, \citenamefont
  {Fitzsimons},\ and\ \citenamefont
  {Lovett}}]{Close:etal:2011:Spin_State_Amplification}%
  \BibitemOpen
  \bibfield  {author} {\bibinfo {author} {\bibfnamefont {T.}~\bibnamefont
  {Close}}, \bibinfo {author} {\bibfnamefont {F.}~\bibnamefont {Fadugba}},
  \bibinfo {author} {\bibfnamefont {S.~C.}\ \bibnamefont {Benjamin}}, \bibinfo
  {author} {\bibfnamefont {J.}~\bibnamefont {Fitzsimons}}, \ and\ \bibinfo
  {author} {\bibfnamefont {B.~W.}\ \bibnamefont {Lovett}},\ }\href {\doibase
  10.1103/PhysRevLett.106.167204} {\bibfield  {journal} {\bibinfo  {journal}
  {Phys. Rev. Lett.}\ }\textbf {\bibinfo {volume} {106}},\ \bibinfo {pages}
  {167204} (\bibinfo {year} {2011})}\BibitemShut {NoStop}%
\bibitem [{\citenamefont {Sturm}\ \emph {et~al.}(2011)\citenamefont {Sturm},
  \citenamefont {Wagner}, \citenamefont {Schabinger}, \citenamefont {Zatorski},
  \citenamefont {Harman}, \citenamefont {Quint}, \citenamefont {Werth},
  \citenamefont {Keitel},\ and\ \citenamefont
  {Blaum}}]{Sturm:etal:2011:g_factor}%
  \BibitemOpen
  \bibfield  {author} {\bibinfo {author} {\bibfnamefont {S.}~\bibnamefont
  {Sturm}}, \bibinfo {author} {\bibfnamefont {A.}~\bibnamefont {Wagner}},
  \bibinfo {author} {\bibfnamefont {B.}~\bibnamefont {Schabinger}}, \bibinfo
  {author} {\bibfnamefont {J.}~\bibnamefont {Zatorski}}, \bibinfo {author}
  {\bibfnamefont {Z.}~\bibnamefont {Harman}}, \bibinfo {author} {\bibfnamefont
  {W.}~\bibnamefont {Quint}}, \bibinfo {author} {\bibfnamefont
  {G.}~\bibnamefont {Werth}}, \bibinfo {author} {\bibfnamefont {C.~H.}\
  \bibnamefont {Keitel}}, \ and\ \bibinfo {author} {\bibfnamefont
  {K.}~\bibnamefont {Blaum}},\ }\href {\doibase 10.1103/PhysRevLett.107.023002}
  {\bibfield  {journal} {\bibinfo  {journal} {Phys. Rev. Lett.}\ }\textbf
  {\bibinfo {volume} {107}},\ \bibinfo {pages} {023002} (\bibinfo {year}
  {2011})}\BibitemShut {NoStop}%
\bibitem [{\citenamefont {DiSciacca}\ and\ \citenamefont
  {Gabrielse}(2012)}]{DiSciacca:2012:Direct_Measurement}%
  \BibitemOpen
  \bibfield  {author} {\bibinfo {author} {\bibfnamefont {J.}~\bibnamefont
  {DiSciacca}}\ and\ \bibinfo {author} {\bibfnamefont {G.}~\bibnamefont
  {Gabrielse}},\ }\href {\doibase 10.1103/PhysRevLett.108.153001} {\bibfield
  {journal} {\bibinfo  {journal} {Phys. Rev. Lett.}\ }\textbf {\bibinfo
  {volume} {108}},\ \bibinfo {pages} {153001} (\bibinfo {year}
  {2012})}\BibitemShut {NoStop}%
\bibitem [{\citenamefont {Mott}(1929)}]{Mott:1929:Scattering}%
  \BibitemOpen
  \bibfield  {author} {\bibinfo {author} {\bibfnamefont {N.~F.}\ \bibnamefont
  {Mott}},\ }\href {\doibase 10.1098/rspa.1929.0127} {\bibfield  {journal}
  {\bibinfo  {journal} {Proc. R. Soc. Lond. A}\ }\textbf {\bibinfo {volume}
  {124}},\ \bibinfo {pages} {425} (\bibinfo {year} {1929})}\BibitemShut
  {NoStop}%
\bibitem [{\citenamefont {Peres}\ and\ \citenamefont
  {Terno}(2004)}]{Peres:Terno:2004:QI_RT}%
  \BibitemOpen
  \bibfield  {author} {\bibinfo {author} {\bibfnamefont {A.}~\bibnamefont
  {Peres}}\ and\ \bibinfo {author} {\bibfnamefont {D.~R.}\ \bibnamefont
  {Terno}},\ }\href {\doibase 10.1103/RevModPhys.76.93} {\bibfield  {journal}
  {\bibinfo  {journal} {Rev. Mod. Phys.}\ }\textbf {\bibinfo {volume} {76}},\
  \bibinfo {pages} {93} (\bibinfo {year} {2004})}\BibitemShut {NoStop}%
\bibitem [{\citenamefont {Kim}\ and\ \citenamefont
  {Son}(2005)}]{Kim:etal:2005:Bell}%
  \BibitemOpen
  \bibfield  {author} {\bibinfo {author} {\bibfnamefont {W.~T.}\ \bibnamefont
  {Kim}}\ and\ \bibinfo {author} {\bibfnamefont {E.~J.}\ \bibnamefont {Son}},\
  }\href {\doibase 10.1103/PhysRevA.71.014102} {\bibfield  {journal} {\bibinfo
  {journal} {Phys. Rev. A}\ }\textbf {\bibinfo {volume} {71}},\ \bibinfo
  {pages} {014102} (\bibinfo {year} {2005})}\BibitemShut {NoStop}%
\bibitem [{\citenamefont
  {Wiesendanger}(2009)}]{Wiesendanger:2009:Spin_mapping}%
  \BibitemOpen
  \bibfield  {author} {\bibinfo {author} {\bibfnamefont {R.}~\bibnamefont
  {Wiesendanger}},\ }\href {\doibase 10.1103/RevModPhys.81.1495} {\bibfield
  {journal} {\bibinfo  {journal} {Rev. Mod. Phys.}\ }\textbf {\bibinfo {volume}
  {81}},\ \bibinfo {pages} {1495} (\bibinfo {year} {2009})}\BibitemShut
  {NoStop}%
\bibitem [{\citenamefont {Friis}\ \emph {et~al.}(2010)\citenamefont {Friis},
  \citenamefont {Bertlmann}, \citenamefont {Huber},\ and\ \citenamefont
  {Hiesmayr}}]{Friis:etal:2010}%
  \BibitemOpen
  \bibfield  {author} {\bibinfo {author} {\bibfnamefont {N.}~\bibnamefont
  {Friis}}, \bibinfo {author} {\bibfnamefont {R.~A.}\ \bibnamefont
  {Bertlmann}}, \bibinfo {author} {\bibfnamefont {M.}~\bibnamefont {Huber}}, \
  and\ \bibinfo {author} {\bibfnamefont {B.~C.}\ \bibnamefont {Hiesmayr}},\
  }\href {\doibase 10.1103/PhysRevA.81.042114} {\bibfield  {journal} {\bibinfo
  {journal} {Phys. Rev. A}\ }\textbf {\bibinfo {volume} {81}},\ \bibinfo
  {pages} {042114} (\bibinfo {year} {2010})}\BibitemShut {NoStop}%
\bibitem [{\citenamefont {Petersson}\ \emph {et~al.}(2012)\citenamefont
  {Petersson}, \citenamefont {McFaul}, \citenamefont {Schroer}, \citenamefont
  {Jung}, \citenamefont {Taylor}, \citenamefont {Houck},\ and\ \citenamefont
  {Petta}}]{Petersson:2010:Circuit_qed}%
  \BibitemOpen
  \bibfield  {author} {\bibinfo {author} {\bibfnamefont {K.~D.}\ \bibnamefont
  {Petersson}}, \bibinfo {author} {\bibfnamefont {L.~W.}\ \bibnamefont
  {McFaul}}, \bibinfo {author} {\bibfnamefont {M.~D.}\ \bibnamefont {Schroer}},
  \bibinfo {author} {\bibfnamefont {M.}~\bibnamefont {Jung}}, \bibinfo {author}
  {\bibfnamefont {J.~M.}\ \bibnamefont {Taylor}}, \bibinfo {author}
  {\bibfnamefont {A.~A.}\ \bibnamefont {Houck}}, \ and\ \bibinfo {author}
  {\bibfnamefont {J.~R.}\ \bibnamefont {Petta}},\ }\href {\doibase
  10.1038/nature11559} {\bibfield  {journal} {\bibinfo  {journal} {Nature}\
  }\textbf {\bibinfo {volume} {490}},\ \bibinfo {pages} {380} (\bibinfo {year}
  {2012})}\BibitemShut {NoStop}%
\bibitem [{\citenamefont {Saldanha}\ and\ \citenamefont
  {Vedral}(2012)}]{Saldanha:Vedral:2012:Wigner}%
  \BibitemOpen
  \bibfield  {author} {\bibinfo {author} {\bibfnamefont {P.~L.}\ \bibnamefont
  {Saldanha}}\ and\ \bibinfo {author} {\bibfnamefont {V.}~\bibnamefont
  {Vedral}},\ }\href {\doibase 10.1088/1367-2630/14/2/023041} {\bibfield
  {journal} {\bibinfo  {journal} {New J. Phys.}\ }\textbf {\bibinfo {volume}
  {14}},\ \bibinfo {pages} {023041} (\bibinfo {year} {2012})}\BibitemShut
  {NoStop}%
\bibitem [{\citenamefont {Awschalom}\ \emph {et~al.}(2013)\citenamefont
  {Awschalom}, \citenamefont {Bassett}, \citenamefont {Dzurak}, \citenamefont
  {Hu},\ and\ \citenamefont {Petta}}]{Awschalom:etal:2013:Quantum_Spintronics}%
  \BibitemOpen
  \bibfield  {author} {\bibinfo {author} {\bibfnamefont {D.~D.}\ \bibnamefont
  {Awschalom}}, \bibinfo {author} {\bibfnamefont {L.~C.}\ \bibnamefont
  {Bassett}}, \bibinfo {author} {\bibfnamefont {A.~S.}\ \bibnamefont {Dzurak}},
  \bibinfo {author} {\bibfnamefont {E.~L.}\ \bibnamefont {Hu}}, \ and\ \bibinfo
  {author} {\bibfnamefont {J.~R.}\ \bibnamefont {Petta}},\ }\href {\doibase
  10.1126/science.1231364} {\bibfield  {journal} {\bibinfo  {journal}
  {Science}\ }\textbf {\bibinfo {volume} {339}},\ \bibinfo {pages} {1174}
  (\bibinfo {year} {2013})}\BibitemShut {NoStop}%
\bibitem [{\citenamefont {Abanin}\ \emph {et~al.}(2011)\citenamefont {Abanin},
  \citenamefont {Gorbachev}, \citenamefont {Novoselov}, \citenamefont {Geim},\
  and\ \citenamefont {Levitov}}]{Abanin:2011:Giant_Spin-Hall_Effect}%
  \BibitemOpen
  \bibfield  {author} {\bibinfo {author} {\bibfnamefont {D.~A.}\ \bibnamefont
  {Abanin}}, \bibinfo {author} {\bibfnamefont {R.~V.}\ \bibnamefont
  {Gorbachev}}, \bibinfo {author} {\bibfnamefont {K.~S.}\ \bibnamefont
  {Novoselov}}, \bibinfo {author} {\bibfnamefont {A.~K.}\ \bibnamefont {Geim}},
  \ and\ \bibinfo {author} {\bibfnamefont {L.~S.}\ \bibnamefont {Levitov}},\
  }\href {\doibase 10.1103/PhysRevLett.107.096601} {\bibfield  {journal}
  {\bibinfo  {journal} {Phys. Rev. Lett.}\ }\textbf {\bibinfo {volume} {107}},\
  \bibinfo {pages} {096601} (\bibinfo {year} {2011})}\BibitemShut {NoStop}%
\bibitem [{\citenamefont {Mecklenburg}\ and\ \citenamefont
  {Regan}(2011)}]{Mecklenburg:Regan:2011:Spin}%
  \BibitemOpen
  \bibfield  {author} {\bibinfo {author} {\bibfnamefont {M.}~\bibnamefont
  {Mecklenburg}}\ and\ \bibinfo {author} {\bibfnamefont {B.~C.}\ \bibnamefont
  {Regan}},\ }\href {\doibase 10.1103/PhysRevLett.106.116803} {\bibfield
  {journal} {\bibinfo  {journal} {Phys. Rev. Lett.}\ }\textbf {\bibinfo
  {volume} {106}},\ \bibinfo {pages} {116803} (\bibinfo {year}
  {2011})}\BibitemShut {NoStop}%
\bibitem [{\citenamefont {G\"uttinger}\ \emph {et~al.}(2010)\citenamefont
  {G\"uttinger}, \citenamefont {Frey}, \citenamefont {Stampfer}, \citenamefont
  {Ihn},\ and\ \citenamefont {Ensslin}}]{Guttinger:2010:Spin_States}%
  \BibitemOpen
  \bibfield  {author} {\bibinfo {author} {\bibfnamefont {J.}~\bibnamefont
  {G\"uttinger}}, \bibinfo {author} {\bibfnamefont {T.}~\bibnamefont {Frey}},
  \bibinfo {author} {\bibfnamefont {C.}~\bibnamefont {Stampfer}}, \bibinfo
  {author} {\bibfnamefont {T.}~\bibnamefont {Ihn}}, \ and\ \bibinfo {author}
  {\bibfnamefont {K.}~\bibnamefont {Ensslin}},\ }\href {\doibase
  10.1103/PhysRevLett.105.116801} {\bibfield  {journal} {\bibinfo  {journal}
  {Phys. Rev. Lett.}\ }\textbf {\bibinfo {volume} {105}},\ \bibinfo {pages}
  {116801} (\bibinfo {year} {2010})}\BibitemShut {NoStop}%
\bibitem [{\citenamefont {{Di Piazza}}\ \emph {et~al.}(2012)\citenamefont {{Di
  Piazza}}, \citenamefont {M\"uller}, \citenamefont {Hatsagortsyan},\ and\
  \citenamefont {Keitel}}]{DiPiazza:etal:2012:high-intensity}%
  \BibitemOpen
  \bibfield  {author} {\bibinfo {author} {\bibfnamefont {A.}~\bibnamefont {{Di
  Piazza}}}, \bibinfo {author} {\bibfnamefont {C.}~\bibnamefont {M\"uller}},
  \bibinfo {author} {\bibfnamefont {K.~Z.}\ \bibnamefont {Hatsagortsyan}}, \
  and\ \bibinfo {author} {\bibfnamefont {C.~H.}\ \bibnamefont {Keitel}},\
  }\href {\doibase 10.1103/RevModPhys.84.1177} {\bibfield  {journal} {\bibinfo
  {journal} {Rev. Mod. Phys.}\ }\textbf {\bibinfo {volume} {84}},\ \bibinfo
  {pages} {1177} (\bibinfo {year} {2012})}\BibitemShut {NoStop}%
\bibitem [{\citenamefont {Ahrens}\ \emph {et~al.}(2012)\citenamefont {Ahrens},
  \citenamefont {Bauke}, \citenamefont {Keitel},\ and\ \citenamefont
  {M\"uller}}]{Ahrens:etal:2012:spin_dynamics}%
  \BibitemOpen
  \bibfield  {author} {\bibinfo {author} {\bibfnamefont {S.}~\bibnamefont
  {Ahrens}}, \bibinfo {author} {\bibfnamefont {H.}~\bibnamefont {Bauke}},
  \bibinfo {author} {\bibfnamefont {C.~H.}\ \bibnamefont {Keitel}}, \ and\
  \bibinfo {author} {\bibfnamefont {C.}~\bibnamefont {M\"uller}},\ }\href
  {\doibase 10.1103/PhysRevLett.109.043601} {\bibfield  {journal} {\bibinfo
  {journal} {Phys. Rev. Lett.}\ }\textbf {\bibinfo {volume} {109}},\ \bibinfo
  {pages} {043601} (\bibinfo {year} {2012})}\BibitemShut {NoStop}%
\bibitem [{\citenamefont {Ahrens}\ \emph {et~al.}(2014)\citenamefont {Ahrens},
  \citenamefont {Bauke}, \citenamefont {Keitel},\ and\ \citenamefont
  {Grobe}}]{Ahrens:etal:2014:Electron-spin_dynamics_induced_by_photon_spins}%
  \BibitemOpen
  \bibfield  {author} {\bibinfo {author} {\bibfnamefont {S.}~\bibnamefont
  {Ahrens}}, \bibinfo {author} {\bibfnamefont {H.}~\bibnamefont {Bauke}},
  \bibinfo {author} {\bibfnamefont {C.~H.}\ \bibnamefont {Keitel}}, \ and\
  \bibinfo {author} {\bibfnamefont {R.}~\bibnamefont {Grobe}},\ }\href@noop {}
  {\enquote {\bibinfo {title} {Electron-spin dynamics induced by photon
  spins},}\ } (\bibinfo {year} {2014}),\ \Eprint {arXiv/1401.5976}
  {arXiv:1401.5976} \BibitemShut {NoStop}%
\bibitem [{\citenamefont {Newton}\ and\ \citenamefont
  {Wigner}(1949)}]{Newton:Wigner:1949:Localized_States}%
  \BibitemOpen
  \bibfield  {author} {\bibinfo {author} {\bibfnamefont {T.~D.}\ \bibnamefont
  {Newton}}\ and\ \bibinfo {author} {\bibfnamefont {E.~P.}\ \bibnamefont
  {Wigner}},\ }\href {\doibase 10.1103/RevModPhys.21.400} {\bibfield  {journal}
  {\bibinfo  {journal} {Rev. Mod. Phys.}\ }\textbf {\bibinfo {volume} {21}},\
  \bibinfo {pages} {400} (\bibinfo {year} {1949})}\BibitemShut {NoStop}%
\bibitem [{\citenamefont {Jordan}\ and\ \citenamefont
  {Mukunda}(1963)}]{Jordan:Mukunda:1963:Lorentz-Covariant_Position_Operators}%
  \BibitemOpen
  \bibfield  {author} {\bibinfo {author} {\bibfnamefont {T.~F.}\ \bibnamefont
  {Jordan}}\ and\ \bibinfo {author} {\bibfnamefont {N.}~\bibnamefont
  {Mukunda}},\ }\href {\doibase 10.1103/PhysRev.132.1842} {\bibfield  {journal}
  {\bibinfo  {journal} {Phys. Rev.}\ }\textbf {\bibinfo {volume} {132}},\
  \bibinfo {pages} {1842} (\bibinfo {year} {1963})}\BibitemShut {NoStop}%
\bibitem [{\citenamefont {O'Connell}\ and\ \citenamefont
  {Wigner}(1978)}]{OConnell:Wigner:1978:Position_operators}%
  \BibitemOpen
  \bibfield  {author} {\bibinfo {author} {\bibfnamefont {R.}~\bibnamefont
  {O'Connell}}\ and\ \bibinfo {author} {\bibfnamefont {E.}~\bibnamefont
  {Wigner}},\ }\href {\doibase 10.1016/0375-9601(78)90317-1} {\bibfield
  {journal} {\bibinfo  {journal} {Phys. Lett. A}\ }\textbf {\bibinfo {volume}
  {67}},\ \bibinfo {pages} {319} (\bibinfo {year} {1978})}\BibitemShut
  {NoStop}%
\bibitem [{\citenamefont {Bethe}\ and\ \citenamefont
  {Salpeter}(2008)}]{Bethe:Salpeter:2008:Atoms}%
  \BibitemOpen
  \bibfield  {author} {\bibinfo {author} {\bibfnamefont {H.~A.}\ \bibnamefont
  {Bethe}}\ and\ \bibinfo {author} {\bibfnamefont {E.~E.}\ \bibnamefont
  {Salpeter}},\ }\href@noop {} {\emph {\bibinfo {title} {Quantum Mechanics of
  One- and Two-Electron Atoms}}}\ (\bibinfo  {publisher} {Dover},\ \bibinfo
  {address} {Mineola},\ \bibinfo {year} {2008})\BibitemShut {NoStop}%
\bibitem [{\citenamefont {Thaller}(2000)}]{Thaller:2005:vis_QM_II}%
  \BibitemOpen
  \bibfield  {author} {\bibinfo {author} {\bibfnamefont {B.}~\bibnamefont
  {Thaller}},\ }\href@noop {} {\emph {\bibinfo {title} {Advanced Visual Quantum
  Mechanics}}}\ (\bibinfo  {publisher} {Springer},\ \bibinfo {address}
  {Heidelberg, New York},\ \bibinfo {year} {2000})\BibitemShut {NoStop}%
\bibitem [{\citenamefont {Hill}\ and\ \citenamefont
  {Landshoff}(1938)}]{Hill:Landshoff:1938}%
  \BibitemOpen
  \bibfield  {author} {\bibinfo {author} {\bibfnamefont {E.~L.}\ \bibnamefont
  {Hill}}\ and\ \bibinfo {author} {\bibfnamefont {R.}~\bibnamefont
  {Landshoff}},\ }\href {\doibase 10.1103/RevModPhys.10.87} {\bibfield
  {journal} {\bibinfo  {journal} {Rev. Mod. Phys.}\ }\textbf {\bibinfo {volume}
  {10}},\ \bibinfo {pages} {87} (\bibinfo {year} {1938})}\BibitemShut {NoStop}%
\bibitem [{\citenamefont {Dirac}(1971)}]{Dirac:1971:Positive-Energy}%
  \BibitemOpen
  \bibfield  {author} {\bibinfo {author} {\bibfnamefont {P.~A.~M.}\
  \bibnamefont {Dirac}},\ }\href {\doibase 10.1098/rspa.1971.0077} {\bibfield
  {journal} {\bibinfo  {journal} {Proc. R. Soc. London, Ser. A}\ }\textbf
  {\bibinfo {volume} {322}},\ \bibinfo {pages} {435} (\bibinfo {year}
  {1971})}\BibitemShut {NoStop}%
\bibitem [{\citenamefont {Ohanian}(1986)}]{Ohanian:1986:spin}%
  \BibitemOpen
  \bibfield  {author} {\bibinfo {author} {\bibfnamefont {H.~C.}\ \bibnamefont
  {Ohanian}},\ }\href {\doibase 10.1119/1.14580} {\bibfield  {journal}
  {\bibinfo  {journal} {Amer. J. Phys.}\ }\textbf {\bibinfo {volume} {54}},\
  \bibinfo {pages} {500} (\bibinfo {year} {1986})}\BibitemShut {NoStop}%
\bibitem [{\citenamefont {Foldy}\ and\ \citenamefont
  {Wouthuysen}(1950)}]{Foldy:Wouthuysen:1950:Non-Relativistic_Limit}%
  \BibitemOpen
  \bibfield  {author} {\bibinfo {author} {\bibfnamefont {L.~L.}\ \bibnamefont
  {Foldy}}\ and\ \bibinfo {author} {\bibfnamefont {S.~A.}\ \bibnamefont
  {Wouthuysen}},\ }\href {\doibase 10.1103/PhysRev.78.29} {\bibfield  {journal}
  {\bibinfo  {journal} {Phys. Rev.}\ }\textbf {\bibinfo {volume} {78}},\
  \bibinfo {pages} {29} (\bibinfo {year} {1950})}\BibitemShut {NoStop}%
\bibitem [{\citenamefont {{de Vries}}(1970)}]{deVries:1970:FW_Transformation}%
  \BibitemOpen
  \bibfield  {author} {\bibinfo {author} {\bibfnamefont {E.}~\bibnamefont {{de
  Vries}}},\ }\href {\doibase 10.1002/prop.19700180402} {\bibfield  {journal}
  {\bibinfo  {journal} {Fortschr. Phys.}\ }\textbf {\bibinfo {volume} {18}},\
  \bibinfo {pages} {149} (\bibinfo {year} {1970})}\BibitemShut {NoStop}%
\bibitem [{\citenamefont {Costella}\ and\ \citenamefont
  {McKellar}(1995)}]{Costella:McKellar:1995:Foldy-Wouthuysen}%
  \BibitemOpen
  \bibfield  {author} {\bibinfo {author} {\bibfnamefont {J.~P.}\ \bibnamefont
  {Costella}}\ and\ \bibinfo {author} {\bibfnamefont {B.~H.~J.}\ \bibnamefont
  {McKellar}},\ }\href {\doibase 10.1119/1.18017} {\bibfield  {journal}
  {\bibinfo  {journal} {Amer. J. Phys.}\ }\textbf {\bibinfo {volume} {63}},\
  \bibinfo {pages} {1119} (\bibinfo {year} {1995})}\BibitemShut {NoStop}%
\bibitem [{\citenamefont {Schweber}(2005)}]{Schweber:2005:qft}%
  \BibitemOpen
  \bibfield  {author} {\bibinfo {author} {\bibfnamefont {S.~S.}\ \bibnamefont
  {Schweber}},\ }\href@noop {} {\emph {\bibinfo {title} {An introduction to
  relativistic quantum field theory}}}\ (\bibinfo  {publisher} {Dover},\
  \bibinfo {address} {Mineola},\ \bibinfo {year} {2005})\BibitemShut {NoStop}%
\bibitem [{\citenamefont {Caban}\ \emph {et~al.}(2013)\citenamefont {Caban},
  \citenamefont {Rembieli\'nski},\ and\ \citenamefont {{W\l
  odarczyk}}}]{Caban:etal:2013:A_spin_observable}%
  \BibitemOpen
  \bibfield  {author} {\bibinfo {author} {\bibfnamefont {P.}~\bibnamefont
  {Caban}}, \bibinfo {author} {\bibfnamefont {J.}~\bibnamefont
  {Rembieli\'nski}}, \ and\ \bibinfo {author} {\bibfnamefont {M.}~\bibnamefont
  {{W\l odarczyk}}},\ }\href {\doibase 10.1016/j.aop.2012.12.001} {\bibfield
  {journal} {\bibinfo  {journal} {Ann. Phys.}\ }\textbf {\bibinfo {volume}
  {330}},\ \bibinfo {pages} {263} (\bibinfo {year} {2013})}\BibitemShut
  {NoStop}%
\bibitem [{\citenamefont {Czachor}(1997)}]{Czachor:1997:relativistic_EPR}%
  \BibitemOpen
  \bibfield  {author} {\bibinfo {author} {\bibfnamefont {M.}~\bibnamefont
  {Czachor}},\ }\href {\doibase 10.1103/PhysRevA.55.72} {\bibfield  {journal}
  {\bibinfo  {journal} {Phys. Rev. A}\ }\textbf {\bibinfo {volume} {55}},\
  \bibinfo {pages} {72} (\bibinfo {year} {1997})}\BibitemShut {NoStop}%
\bibitem [{\citenamefont {Hilgevoord}\ and\ \citenamefont
  {Wouthuysen}(1963)}]{Hilgevoord:Wouthuysen:1963:Dirac_spin}%
  \BibitemOpen
  \bibfield  {author} {\bibinfo {author} {\bibfnamefont {J.}~\bibnamefont
  {Hilgevoord}}\ and\ \bibinfo {author} {\bibfnamefont {S.~A.}\ \bibnamefont
  {Wouthuysen}},\ }\href {\doibase 10.1016/0029-5582(63)90246-3} {\bibfield
  {journal} {\bibinfo  {journal} {Nucl. Phys.}\ }\textbf {\bibinfo {volume}
  {40}},\ \bibinfo {pages} {1} (\bibinfo {year} {1963})}\BibitemShut {NoStop}%
\bibitem [{\citenamefont {Wightman}(1960)}]{Wightman:1960}%
  \BibitemOpen
  \bibfield  {author} {\bibinfo {author} {\bibfnamefont {A.~S.}\ \bibnamefont
  {Wightman}},\ }\enquote {\bibinfo {title} {L'invariance dans la mecanique
  quantique relativiste},}\ \ (\bibinfo  {publisher} {Herman},\ \bibinfo {year}
  {1960})\ Chap.\ \bibinfo {chapter} {V: Relations de dispersion et particules
  {\'e}l{\'e}mentaires}\BibitemShut {NoStop}%
\bibitem [{\citenamefont {Bargmann}\ \emph {et~al.}(1935)\citenamefont
  {Bargmann}, \citenamefont {Michel},\ and\ \citenamefont
  {Telegdi}}]{Bargmann:etal:1935:Precession}%
  \BibitemOpen
  \bibfield  {author} {\bibinfo {author} {\bibfnamefont {V.}~\bibnamefont
  {Bargmann}}, \bibinfo {author} {\bibfnamefont {L.}~\bibnamefont {Michel}}, \
  and\ \bibinfo {author} {\bibfnamefont {V.~L.}\ \bibnamefont {Telegdi}},\
  }\href {\doibase 10.1103/PhysRevLett.2.435} {\bibfield  {journal} {\bibinfo
  {journal} {Phys. Rev. Lett.}\ }\textbf {\bibinfo {volume} {2}},\ \bibinfo
  {pages} {435} (\bibinfo {year} {1935})}\BibitemShut {NoStop}%
\bibitem [{\citenamefont
  {Chakrabarti}(1963)}]{Chakrabarti:1963:Canonical_Form}%
  \BibitemOpen
  \bibfield  {author} {\bibinfo {author} {\bibfnamefont {A.}~\bibnamefont
  {Chakrabarti}},\ }\href {\doibase 10.1063/1.1703892} {\bibfield  {journal}
  {\bibinfo  {journal} {J. Math. Phys.}\ }\textbf {\bibinfo {volume} {4}},\
  \bibinfo {pages} {1215} (\bibinfo {year} {1963})}\BibitemShut {NoStop}%
\bibitem [{\citenamefont
  {G\"ursey}(1965{\natexlab{a}})}]{Gursey:1965:Equivalent}%
  \BibitemOpen
  \bibfield  {author} {\bibinfo {author} {\bibfnamefont {F.}~\bibnamefont
  {G\"ursey}},\ }\href {\doibase 10.1016/0031-9163(65)90226-X} {\bibfield
  {journal} {\bibinfo  {journal} {Phys. Lett.}\ }\textbf {\bibinfo {volume}
  {14}},\ \bibinfo {pages} {330} (\bibinfo {year}
  {1965}{\natexlab{a}})}\BibitemShut {NoStop}%
\bibitem [{\citenamefont {G\"ursey}(1965{\natexlab{b}})}]{Gursey:1965}%
  \BibitemOpen
  \bibfield  {author} {\bibinfo {author} {\bibfnamefont {F.}~\bibnamefont
  {G\"ursey}},\ }in\ \href@noop {} {\emph {\bibinfo {booktitle} {High energy
  physics}}}\ (\bibinfo  {publisher} {Gordon and Breach},\ \bibinfo {year}
  {1965})\ pp.\ \bibinfo {pages} {53--88}\BibitemShut {NoStop}%
\bibitem [{\citenamefont
  {G\"ursey}(1965{\natexlab{c}})}]{Gursey:1965:Equivalent_formulations}%
  \BibitemOpen
  \bibfield  {author} {\bibinfo {author} {\bibfnamefont {F.}~\bibnamefont
  {G\"ursey}},\ }\href {\doibase 10.1016/0031-9163(65)90226-X} {\bibfield
  {journal} {\bibinfo  {journal} {Phys. Lett.}\ }\textbf {\bibinfo {volume}
  {14}},\ \bibinfo {pages} {330} (\bibinfo {year}
  {1965}{\natexlab{c}})}\BibitemShut {NoStop}%
\bibitem [{\citenamefont {Kirsch}\ \emph {et~al.}(2001)\citenamefont {Kirsch},
  \citenamefont {Ryder},\ and\ \citenamefont
  {Hehl}}]{Kirsch:Ryder:Hehl:2001:Gordon_decompositions}%
  \BibitemOpen
  \bibfield  {author} {\bibinfo {author} {\bibfnamefont {I.}~\bibnamefont
  {Kirsch}}, \bibinfo {author} {\bibfnamefont {L.~H.}\ \bibnamefont {Ryder}}, \
  and\ \bibinfo {author} {\bibfnamefont {F.~W.}\ \bibnamefont {Hehl}},\
  }\href@noop {} {\enquote {\bibinfo {title} {The {Gordon} decompositions of
  the inertial currents of the {Dirac} electron correspond to a
  {Foldy-Wouthuysen} transformation},}\ } (\bibinfo {year} {2001}),\ \Eprint
  {arXiv/hep-th/0102102} {arXiv:hep-th/0102102} \BibitemShut {NoStop}%
\bibitem [{\citenamefont {Ryder}(1999)}]{Ryder:1999:Relativistic_Spin}%
  \BibitemOpen
  \bibfield  {author} {\bibinfo {author} {\bibfnamefont {L.~H.}\ \bibnamefont
  {Ryder}},\ }\href {\doibase 10.1023/A:1026669717679} {\bibfield  {journal}
  {\bibinfo  {journal} {General Relativity and Gravitation}\ }\textbf {\bibinfo
  {volume} {31}},\ \bibinfo {pages} {775} (\bibinfo {year} {1999})}\BibitemShut
  {NoStop}%
\bibitem [{\citenamefont {Choi}(2013)}]{Choi:2013:Relativistic_spin_operator}%
  \BibitemOpen
  \bibfield  {author} {\bibinfo {author} {\bibfnamefont {T.}~\bibnamefont
  {Choi}},\ }\href {\doibase 10.3938/jkps.62.1085} {\bibfield  {journal}
  {\bibinfo  {journal} {J. Korean Phys. Soc.}\ }\textbf {\bibinfo {volume}
  {62}},\ \bibinfo {pages} {1085} (\bibinfo {year} {2013})}\BibitemShut
  {NoStop}%
\bibitem [{\citenamefont {Pryce}(1935)}]{Pryce:1935:Commuting_Co-Ordinates}%
  \BibitemOpen
  \bibfield  {author} {\bibinfo {author} {\bibfnamefont {M.~H.~L.}\
  \bibnamefont {Pryce}},\ }\href {\doibase 10.1098/rspa.1935.0094} {\bibfield
  {journal} {\bibinfo  {journal} {Proc. R. Soc. London, Ser. A}\ }\textbf
  {\bibinfo {volume} {150}},\ \bibinfo {pages} {166} (\bibinfo {year}
  {1935})}\BibitemShut {NoStop}%
\bibitem [{\citenamefont {Pryce}(1948)}]{Pryce:1948:Mass-Centre}%
  \BibitemOpen
  \bibfield  {author} {\bibinfo {author} {\bibfnamefont {M.~H.~L.}\
  \bibnamefont {Pryce}},\ }\href {\doibase 10.1098/rspa.1948.0103} {\bibfield
  {journal} {\bibinfo  {journal} {Proc. R. Soc. London, Ser. A}\ }\textbf
  {\bibinfo {volume} {195}},\ \bibinfo {pages} {62} (\bibinfo {year}
  {1948})}\BibitemShut {NoStop}%
\bibitem [{\citenamefont
  {Macfarlane}(1963)}]{Macfarlane:1963:Relativistic_Two_Particle}%
  \BibitemOpen
  \bibfield  {author} {\bibinfo {author} {\bibfnamefont {A.~J.}\ \bibnamefont
  {Macfarlane}},\ }\href {\doibase 10.1063/1.1703981} {\bibfield  {journal}
  {\bibinfo  {journal} {J. Math. Phys.}\ }\textbf {\bibinfo {volume} {4}},\
  \bibinfo {pages} {490} (\bibinfo {year} {1963})}\BibitemShut {NoStop}%
\bibitem [{\citenamefont {Berg}(1965)}]{Berg:1965:Position_Spin}%
  \BibitemOpen
  \bibfield  {author} {\bibinfo {author} {\bibfnamefont {R.~A.}\ \bibnamefont
  {Berg}},\ }\href {\doibase 10.1063/1.1704260} {\bibfield  {journal} {\bibinfo
   {journal} {J. Math. P.}\ }\textbf {\bibinfo {volume} {6}},\ \bibinfo {pages}
  {34} (\bibinfo {year} {1965})}\BibitemShut {NoStop}%
\bibitem [{\citenamefont {Fradkin}\ and\ \citenamefont
  {Good}(1961)}]{Fradkin:Good:1961:Polarization_Operators}%
  \BibitemOpen
  \bibfield  {author} {\bibinfo {author} {\bibfnamefont {D.~M.}\ \bibnamefont
  {Fradkin}}\ and\ \bibinfo {author} {\bibfnamefont {R.~H.}\ \bibnamefont
  {Good}, \bibfnamefont {Jr.}},\ }\href {\doibase 10.1103/RevModPhys.33.343}
  {\bibfield  {journal} {\bibinfo  {journal} {Rev. Mod. Phys.}\ }\textbf
  {\bibinfo {volume} {33}},\ \bibinfo {pages} {343} (\bibinfo {year}
  {1961})}\BibitemShut {NoStop}%
\bibitem [{\citenamefont {Wigner}(1939)}]{Wigner:1939:Unitary_Representations}%
  \BibitemOpen
  \bibfield  {author} {\bibinfo {author} {\bibfnamefont {E.}~\bibnamefont
  {Wigner}},\ }\href@noop {} {\bibfield  {journal} {\bibinfo  {journal} {Annals
  of Mathematics}\ }\textbf {\bibinfo {volume} {40}},\ \bibinfo {pages} {149}
  (\bibinfo {year} {1939})}\BibitemShut {NoStop}%
\bibitem [{\citenamefont {Bargmann}\ and\ \citenamefont
  {Wigner}(1948)}]{Bargmann:Wigner:1948:Group_Theoretical_Discussion}%
  \BibitemOpen
  \bibfield  {author} {\bibinfo {author} {\bibfnamefont {V.}~\bibnamefont
  {Bargmann}}\ and\ \bibinfo {author} {\bibfnamefont {E.~P.}\ \bibnamefont
  {Wigner}},\ }\href@noop {} {\bibfield  {journal} {\bibinfo  {journal} {PNAS}\
  }\textbf {\bibinfo {volume} {34}},\ \bibinfo {pages} {211} (\bibinfo {year}
  {1948})}\BibitemShut {NoStop}%
\bibitem [{\citenamefont {Wigner}(1957)}]{Wigner:1957:Relativistic_Invariance}%
  \BibitemOpen
  \bibfield  {author} {\bibinfo {author} {\bibfnamefont {E.~P.}\ \bibnamefont
  {Wigner}},\ }\href {\doibase 10.1103/RevModPhys.29.255} {\bibfield  {journal}
  {\bibinfo  {journal} {Rev. Mod. Phys.}\ }\textbf {\bibinfo {volume} {29}},\
  \bibinfo {pages} {255} (\bibinfo {year} {1957})}\BibitemShut {NoStop}%
\bibitem [{\citenamefont {Sakurai}\ and\ \citenamefont
  {Napolitano}(2010)}]{Sakurai:Napolitano:2010:Modern_QM}%
  \BibitemOpen
  \bibfield  {author} {\bibinfo {author} {\bibfnamefont {J.~J.}\ \bibnamefont
  {Sakurai}}\ and\ \bibinfo {author} {\bibfnamefont {J.}~\bibnamefont
  {Napolitano}},\ }\href@noop {} {\emph {\bibinfo {title} {Modern Quantum
  Mechanics}}},\ \bibinfo {edition} {2nd}\ ed.\ (\bibinfo  {publisher} {Pearson
  Education},\ \bibinfo {address} {Upper Saddle River},\ \bibinfo {year}
  {2010})\BibitemShut {NoStop}%
\bibitem [{\citenamefont {St{\"o}hlker}\ \emph {et~al.}(1993)\citenamefont
  {St{\"o}hlker}, \citenamefont {Mokler}, \citenamefont {Beckert},
  \citenamefont {Bosch}, \citenamefont {Eickhoff}, \citenamefont {Franzke},
  \citenamefont {Jung}, \citenamefont {Kandler}, \citenamefont {Klepper},
  \citenamefont {Kozhuharov}, \citenamefont {Moshammer}, \citenamefont
  {Nolden}, \citenamefont {Reich}, \citenamefont {Rymuza}, \citenamefont
  {Sp{\"a}dtke}, ,\ and\ \citenamefont
  {Steck}}]{Stohlker:etal:1993:Ground-state_Lamb_shift}%
  \BibitemOpen
  \bibfield  {author} {\bibinfo {author} {\bibfnamefont {T.}~\bibnamefont
  {St{\"o}hlker}}, \bibinfo {author} {\bibfnamefont {P.}~\bibnamefont
  {Mokler}}, \bibinfo {author} {\bibfnamefont {K.}~\bibnamefont {Beckert}},
  \bibinfo {author} {\bibfnamefont {F.}~\bibnamefont {Bosch}}, \bibinfo
  {author} {\bibfnamefont {H.}~\bibnamefont {Eickhoff}}, \bibinfo {author}
  {\bibfnamefont {B.}~\bibnamefont {Franzke}}, \bibinfo {author} {\bibfnamefont
  {M.}~\bibnamefont {Jung}}, \bibinfo {author} {\bibfnamefont {Y.}~\bibnamefont
  {Kandler}}, \bibinfo {author} {\bibfnamefont {O.}~\bibnamefont {Klepper}},
  \bibinfo {author} {\bibfnamefont {C.}~\bibnamefont {Kozhuharov}}, \bibinfo
  {author} {\bibfnamefont {R.}~\bibnamefont {Moshammer}}, \bibinfo {author}
  {\bibfnamefont {F.}~\bibnamefont {Nolden}}, \bibinfo {author} {\bibfnamefont
  {H.}~\bibnamefont {Reich}}, \bibinfo {author} {\bibfnamefont
  {P.}~\bibnamefont {Rymuza}}, \bibinfo {author} {\bibfnamefont
  {P.}~\bibnamefont {Sp{\"a}dtke}}, , \ and\ \bibinfo {author} {\bibfnamefont
  {M.}~\bibnamefont {Steck}},\ }\href {\doibase 10.1103/PhysRevLett.71.2184}
  {\bibfield  {journal} {\bibinfo  {journal} {Phys. Rev. Lett.}\ }\textbf
  {\bibinfo {volume} {71}},\ \bibinfo {pages} {2184} (\bibinfo {year}
  {1993})}\BibitemShut {NoStop}%
\bibitem [{\citenamefont {Robbins}\ \emph {et~al.}(2006)\citenamefont
  {Robbins}, \citenamefont {Beiersdorfer}, \citenamefont {Faenov},
  \citenamefont {Pikuz}, \citenamefont {Thorn}, \citenamefont {Chen},
  \citenamefont {Reed}, \citenamefont {Smith}, \citenamefont {Boyce},
  \citenamefont {Brown}, \citenamefont {Kelley}, \citenamefont {Kilbourne}, ,\
  and\ \citenamefont {Porter}}]{Robbins:etal:2006:Polarization_measurements}%
  \BibitemOpen
  \bibfield  {author} {\bibinfo {author} {\bibfnamefont {D.}~\bibnamefont
  {Robbins}}, \bibinfo {author} {\bibfnamefont {P.}~\bibnamefont
  {Beiersdorfer}}, \bibinfo {author} {\bibfnamefont {A.}~\bibnamefont
  {Faenov}}, \bibinfo {author} {\bibfnamefont {T.}~\bibnamefont {Pikuz}},
  \bibinfo {author} {\bibfnamefont {D.}~\bibnamefont {Thorn}}, \bibinfo
  {author} {\bibfnamefont {H.}~\bibnamefont {Chen}}, \bibinfo {author}
  {\bibfnamefont {K.}~\bibnamefont {Reed}}, \bibinfo {author} {\bibfnamefont
  {A.}~\bibnamefont {Smith}}, \bibinfo {author} {\bibfnamefont
  {K.}~\bibnamefont {Boyce}}, \bibinfo {author} {\bibfnamefont
  {G.}~\bibnamefont {Brown}}, \bibinfo {author} {\bibfnamefont
  {R.}~\bibnamefont {Kelley}}, \bibinfo {author} {\bibfnamefont
  {C.}~\bibnamefont {Kilbourne}}, , \ and\ \bibinfo {author} {\bibfnamefont
  {F.}~\bibnamefont {Porter}},\ }\href {\doibase 10.1103/PhysRevA.74.022713}
  {\bibfield  {journal} {\bibinfo  {journal} {Phys. Rev. A}\ }\textbf {\bibinfo
  {volume} {74}},\ \bibinfo {pages} {022713} (\bibinfo {year}
  {2006})}\BibitemShut {NoStop}%
\bibitem [{\citenamefont {Kluge}\ \emph {et~al.}(2008)\citenamefont {Kluge},
  \citenamefont {Beier}, \citenamefont {Blaum}, \citenamefont {Dahl},
  \citenamefont {Eliseev}, \citenamefont {Herfurth}, \citenamefont {Hofmann},
  \citenamefont {Kester}, \citenamefont {Koszudowski}, \citenamefont
  {Kozhuharov}, \citenamefont {Maero}, \citenamefont {N{\"o}rtersh{\"a}user},
  \citenamefont {Pfister}, \citenamefont {Quint}, \citenamefont {Ratzinger},
  \citenamefont {Schempp}, \citenamefont {Schuch}, \citenamefont
  {St{\"o}hlker}, \citenamefont {Thompson}, \citenamefont {Vogel},
  \citenamefont {Vorobjev}, \citenamefont {Winters},\ and\ \citenamefont
  {Werth}}]{Kluge:etal:2008:HITRAP}%
  \BibitemOpen
  \bibfield  {author} {\bibinfo {author} {\bibfnamefont {H.-J.}\ \bibnamefont
  {Kluge}}, \bibinfo {author} {\bibfnamefont {T.}~\bibnamefont {Beier}},
  \bibinfo {author} {\bibfnamefont {K.}~\bibnamefont {Blaum}}, \bibinfo
  {author} {\bibfnamefont {L.}~\bibnamefont {Dahl}}, \bibinfo {author}
  {\bibfnamefont {S.}~\bibnamefont {Eliseev}}, \bibinfo {author} {\bibfnamefont
  {F.}~\bibnamefont {Herfurth}}, \bibinfo {author} {\bibfnamefont
  {B.}~\bibnamefont {Hofmann}}, \bibinfo {author} {\bibfnamefont
  {O.}~\bibnamefont {Kester}}, \bibinfo {author} {\bibfnamefont
  {S.}~\bibnamefont {Koszudowski}}, \bibinfo {author} {\bibfnamefont
  {C.}~\bibnamefont {Kozhuharov}}, \bibinfo {author} {\bibfnamefont
  {G.}~\bibnamefont {Maero}}, \bibinfo {author} {\bibfnamefont
  {W.}~\bibnamefont {N{\"o}rtersh{\"a}user}}, \bibinfo {author} {\bibfnamefont
  {J.}~\bibnamefont {Pfister}}, \bibinfo {author} {\bibfnamefont
  {W.}~\bibnamefont {Quint}}, \bibinfo {author} {\bibfnamefont
  {U.}~\bibnamefont {Ratzinger}}, \bibinfo {author} {\bibfnamefont
  {A.}~\bibnamefont {Schempp}}, \bibinfo {author} {\bibfnamefont
  {R.}~\bibnamefont {Schuch}}, \bibinfo {author} {\bibfnamefont
  {T.}~\bibnamefont {St{\"o}hlker}}, \bibinfo {author} {\bibfnamefont {R.~C.}\
  \bibnamefont {Thompson}}, \bibinfo {author} {\bibfnamefont {M.}~\bibnamefont
  {Vogel}}, \bibinfo {author} {\bibfnamefont {G.}~\bibnamefont {Vorobjev}},
  \bibinfo {author} {\bibfnamefont {D.~F.~A.}\ \bibnamefont {Winters}}, \ and\
  \bibinfo {author} {\bibfnamefont {G.}~\bibnamefont {Werth}},\ }in\ \href
  {\doibase 10.1016/S0065-3276(07)53007-8} {\emph {\bibinfo {booktitle}
  {Current Trends in Atomic Physics}}},\ \bibinfo {series} {Advances in Quantum
  Chemistry}, Vol.~\bibinfo {volume} {53},\ \bibinfo {editor} {edited by\
  \bibinfo {editor} {\bibfnamefont {S.}~\bibnamefont {Salomonson}}\ and\
  \bibinfo {editor} {\bibfnamefont {E.}~\bibnamefont {Lindroth}}}\ (\bibinfo
  {publisher} {Academic Press},\ \bibinfo {year} {2008})\BibitemShut {NoStop}%
\bibitem [{\citenamefont {Cohen-Tannoudji}\ \emph {et~al.}(1977)\citenamefont
  {Cohen-Tannoudji}, \citenamefont {Diu},\ and\ \citenamefont
  {Lalo\"e}}]{Cohen-Tannoudji:etal:1997:QM_VII}%
  \BibitemOpen
  \bibfield  {author} {\bibinfo {author} {\bibfnamefont {C.}~\bibnamefont
  {Cohen-Tannoudji}}, \bibinfo {author} {\bibfnamefont {B.}~\bibnamefont
  {Diu}}, \ and\ \bibinfo {author} {\bibfnamefont {F.}~\bibnamefont
  {Lalo\"e}},\ }\href@noop {} {\emph {\bibinfo {title} {Quantum Mechanics:
  Volume 2}}}\ (\bibinfo  {publisher} {John Wiley \& Sons},\ \bibinfo {year}
  {1977})\BibitemShut {NoStop}%
\bibitem [{\citenamefont
  {Rubinowicz}(1948)}]{Rubinowicz:1948:Coulomb_Momentum}%
  \BibitemOpen
  \bibfield  {author} {\bibinfo {author} {\bibfnamefont {A.}~\bibnamefont
  {Rubinowicz}},\ }\href {\doibase 10.1103/PhysRev.73.1330} {\bibfield
  {journal} {\bibinfo  {journal} {Phys. Rev.}\ }\textbf {\bibinfo {volume}
  {73}},\ \bibinfo {pages} {1330} (\bibinfo {year} {1948})}\BibitemShut
  {NoStop}%
\end{thebibliography}%

\end{document}